\documentclass[]{emulateapj}

\usepackage{natbib}
\usepackage{amsfonts}
\usepackage{amsmath}
\usepackage{amssymb}
\usepackage{bm}
\usepackage{graphicx}
\usepackage[latin1]{inputenc}
\usepackage{latexsym}
\usepackage{rotating}
\usepackage{bm}             
\usepackage{amsfonts}       

\newcommand{\mb}[1]{\mbox{\boldmath $#1$}}

\def \met  {\mbox{g}}

\newcommand\be{\begin{equation}}
\newcommand\ba{\begin{eqnarray}}
\newcommand\ee{\end{equation}}
\newcommand\ea{\end{eqnarray}}

\newcommand{\cut}{{\textrm{cut}}}
\newcommand{\orb}{{\textrm{orb}}}
\newcommand{\rr}{{\textrm{rr}}}
\newcommand{\Msun}{\textrm{M}_{\odot}}                   
\newcommand{\Mbh}{M_{\bullet}}
\newcommand{\abh}{a_{\bullet}}
\newcommand{\Sbh}{S_{\bullet}}
\newcommand{\Mmw}{M_{\mbox{\tiny MW}}}
\newcommand{\Rbh}{R_{\bullet}}
\newcommand{\Mco}{m}

\begin{document}

\title{Relativistic Effects in Extreme Mass Ratio Gravitational Wave Bursts}


\author{Nicol\'as Yunes, Carlos F.  Sopuerta~\altaffilmark{1,2,3},
  Louis J. Rubbo~\altaffilmark{4}, Kelly
  Holley-Bockelmann~\altaffilmark{5}}

\affil{Institute for Gravitational Physics and Geometry and
  Center for Gravitational Wave Physics, The Pennsylvania State
  University, University Park, PA 16802, USA.}

\altaffiltext{1}{Department of Physics, University of Guelph,
  Guelph, Ontario, Canada N1G 2W1.}

\altaffiltext{2}{Institut de Ci\`encies de l'Espai (CSIC),
  Facultat de Ci\`encies, Campus UAB, Torre C5 parells, E-08193
  Bellaterra (Barcelona), Spain.}

\altaffiltext{3}{Institut d'Estudis Espacials de Catalunya
  (IEEC), Ed. Nexus-201, c/ Gran Capit\`a 2-4, E-08034 Barcelona,
  Spain.}

\altaffiltext{4}{Department of Physics, Coastal Carolina University
  Conway, SC 29528, USA.}

\altaffiltext{5}{Department of Physics and Astronomy, Vanderbilt
  University, Nashville, TN 37235, USA.}


\begin{abstract}
  Extreme mass ratio bursts (EMRBs) have been proposed as a possible
  source for future space-borne gravitational wave detectors, such as
  the Laser Interferometer Space Antenna (LISA).  These events are
  characterized by long-period, nearly-radial orbits of compact
  objects around a central massive black hole. The gravitational
  radiation emitted during such events consists of a short burst,
  corresponding to periapse passage, followed by a longer, silent
  interval. In this paper we investigate the impact of including
  relativistic corrections to the description of the compact object's
  trajectory via a geodesic treatment, as well as including
  higher-order multipole corrections in the waveform calculation.  The
  degree to which the relativistic corrections are important depends
  on the EMRB's orbital parameters.  We find that relativistic EMRBs
  ($v_{\mathrm{max}}/c>0.25$) are not rare and actually account for
  approximately half of the events in our astrophysical model.  The
  relativistic corrections tend to significantly change the waveform
  amplitude and phase relative to a Newtonian description, although
  some of this dephasing could be mimicked by parameter errors.  The
  dephasing over several bursts could be of particular importance not
  only to gravitational wave detection, but also to parameter
  estimation, since it is highly correlated to the spin of the massive
  black hole.  Consequently, we postulate that if a relativistic EMRB
  is detected, such dephasing might be used to probe the relativistic
  character of the massive black hole and obtain information about its
  spin.
\end{abstract}




\maketitle


\section{Introduction} \label{intro}

Low-frequency ($10^{-5} \lesssim f \lesssim 0.1$~Hz) gravitational
wave interferometers, such as the proposed Laser Interferometer Space
Antenna (LISA) \citep{Bender:1998,Danzmann:2003tv,Sumner:2004}, will
open a completely new window to the Universe. Through observations of
low-frequency gravitational radiation we will be able to witness the
inspiral and merger of massive black hole binaries; the inspiral of
compact objects into massive black holes; and millions of
quasi-stationary compact galactic binaries.  Recently, a new source of
low-frequency gravitational radiation has been suggested: extreme mass
ratio bursts (EMRBs) \citep{Rubbo:2006dv,Rubbo:2007}.

EMRBs consist of a stellar-mass compact object (SCO) orbiting a
massive black hole (MBH) of $10^{4-8}~\Msun$ with orbital periods
greater than $T_{\cut} = 3 \times 10^4$~s.  The defining orbital
period cutoff is derived from LISA's lower frequency limit of
$f_{\cut} = 3 \times 10^{-5}$~Hz.  Systems with orbital periods less
than $T_{\cut}$ will radiate continuously inside the LISA band.  Such
continuous systems are more appropriately categorized as extreme mass
ratio inspirals (EMRIs) and have been studied extensively elsewhere:
recent estimations of the event rate are given by \citet{Gair:2004iv}
and \citet{Hopman:2006ha}; a discussion on a possible EMRI background
is given by \citet{Barack:2004b}; accounts on the theoretical
description of EMRIs can be found in the reviews by
\citet{Poisson:2004lr} and \citet{Glampedakis:2005hs}, while the
recent review by \citet{Amaro-Seoane:2007aw} describes the
astrophysical and detection applications.

Although EMRB events are distinct from EMRI events, their evolutionary
track could be connected. In the burst scenario, the SCO orbits the
MBH emitting a beamed burst of gravitational radiation during
pericenter passage.  The emitted radiation carries away energy and
angular momentum from the system so that after multiple pericenter
passages the orbital period decreases, and possibly the system becomes
an EMRI.  However, this evolutionary track is most likely disrupted by
scattering interactions with other stars and/or if the SCO plunges
directly into the central MBH on one of its passages.

The EMRB event rate has recently been investigated using simplified
galactic models and data analysis techniques
\citep{Rubbo:2006dv,Rubbo:2007,Hopman:2006fc}.  Using a density
profile described by an $\eta$-model \citep{Tremaine:1994},
\citet{Rubbo:2006dv,Rubbo:2007} suggested an event rate of
$\sim\!15~{\textrm{yr}}^{-1}$ for events with signal-to-noise ratios
(SNRs) greater than five out to the Virgo cluster.  When mass
segregation and different inner cusp models are considered, the
predicted rate decreases by an order of magnitude
\citep{Hopman:2006fc}.  These preliminary studies were aimed at
understanding if EMRB event rates are interesting for low-frequency
gravitational wave detectors such as LISA.  More work is still needed
to improve the predicted event rate in the context of realistic
galaxies, where the role of non-equilibrium dynamics, anisotropy,
complex star formation histories, substructure, and non-sphericity may
act to change the rate from these fiducial estimates by orders of
magnitude \citep{HB:2006af, Rubbo:2006dv, Rubbo:2007}.

In addition to the astrophysical uncertainties, there are no
investigations of the impact of relativistic corrections to EMRB
dynamics.  All EMRB studies have been carried out in a
\textit{quasi-Newtonian} framework. In this framework, one uses the
Newtonian equations of motion and extracts the gravitational waveforms
by means of the {\em quadrupole formula}.  This approximation ignores
the black hole nature of the central potential, including the black
hole's rotation (spin), and is technically valid only for orbits with
non-relativistic velocities.  However, a considerable number of EMRBs
are characterized by large pericenter velocities ($v_{p} \gtrsim
0.25\,c$) and these \textit{relativistic} EMRBs should produce
gravitational wave signals with larger SNRs, as we will show later.

In this paper, we shall not consider the issue of EMRB event rates,
but instead we shall study the effects of relativistic corrections to
such events.  For extreme-mass-ratio systems a simple way of
introducing relativistic corrections is by using the so-called
\textit{semi-relativistic} approximation introduced by
\citet{Ruffini:1981rs}, and used recently in the context of EMRIs by
\citet{Gair:2005is,Gair:2005hu}.  In this approximation, the MBH and
surrounding area are modeled using the Kerr solution to Einstein's
field equations, which describes a stationary spinning black hole (the
Schwarzschild solution corresponds to the non-spinning case).  The SCO
is considered to be a point-like object (neglecting its own
self-gravity) whose trajectory is described by a geodesic of the Kerr
spacetime.  In other words, relative to previous work, we have
replaced the Newtonian equations of motion by relativistic geodesic
equations of motion.

The relativistic description introduces effects such as orbital
precession and frame dragging, but it does not account for effects due
to the gravitational field induced by the SCO. These effects, for
example, lead to changes in the (geodesic) constants of motion due to
radiation reaction. Even though these effects introduce errors that
scale with the system's mass ratio \citep[e.g.
see][]{Glampedakis:2005hs}, they cannot be neglected for EMRIs. This
is because in the late stages of the EMRI the SCO spends a substantial
fraction of cycles in the strong-field region of the MBH.  On the
other hand, in the case of EMRBs, the SCO sling-shots around the MBH
and its interaction time during pericenter passage is relatively small
($< 10^{5}$~s).  Radiation reaction effects can then be neglected
in EMRBs since the radiation reaction timescale is always much larger
than the period of pericenter passage.

In this paper, we also improve on the semi-relativistic approximation
by using a more precise gravitational wave extraction procedure. The
procedure employed is the multipole-moment wave generation formalism
for slow-motion objects with arbitrarily strong internal gravity
\citep{Thorne:1980rm}.  We consider terms up to the mass-octopole and
current-quadrupole multipoles, thus improving on the mass-quadrupole
analysis of \citet{Rubbo:2006dv,Rubbo:2007} and \citet{Hopman:2006fc}.
Higher multipoles will become important if the system becomes even
more relativistic, but pericenter velocities for EMRBs are typically
small to moderate relative to the speed of light (typically $0.1
\lesssim v_{p}/c \lesssim 0.5$).  Such higher multipolar corrections
were taken into account for EMRIs by~\citet{Babak:2006uv}, but for
those sources the phase evolution must be tracked very accurately,
requiring techniques from black hole perturbation 
theory~\citep{Poisson:2004lr, Glampedakis:2005hs} that we shall not 
consider here.

The study of the relativistic corrections considered in this work
leads to the following conclusions. First, we find that relativistic
effects are significant for approximately $50\%$ of the orbits
contained in the EMRB phase space considered by
\citet{Rubbo:2006dv,Rubbo:2007}.  These relativistic EMRB orbits
differ from their Newtonian counterparts in such a way that the
associated waveforms present a noticeably different structure.  In
particular, we find that there is a dephasing relative to Newtonian
waveforms that is due to precessional effects and depend strongly on
the MBH spin.  These findings show that EMRB events are relativistic
enough that they should be treated accordingly, as was previously
found for EMRIs \citep{Glampedakis:2005hs}.

Second, we find that the corrections to the trajectories affect the
waveforms much more than the corrections in the waveform generation
over several bursts.  For example, for a given relativistic
trajectory, we find that the difference between the SNR of a waveform
obtain from the quadrupole formula to that obtained from the
quadrupole-octopole formula is of the order of $10 \%$ (depending on
the location of the observer.)  On the other hand, using the same
waveform generation formula (quadrupole or quadrupole-octopole), the
difference between the SNR of a Newtonian waveform to that of a Kerr
waveform is of the order of $100 \%$. These findings show that
modeling EMRB waveforms with a quasi-Newtonian treatment might not be
sufficient for certain data analysis applications.

Third, we find that the relativistic corrections accumulate with
multiple bursts and, thus, they may have an important impact in
improving the SNR. It is also conceivable that such corrections might
be important for parameter estimation studies and, perhaps, may be
used to determine or bound the spin of the MBH if a high SNR event is
detected. Along this same lines, if parameters can be determined
accurately enough, it might also be possible to use EMRB measurements
to test deviations from General Relativity. We must note, however,
that changing the orbital parameters in Newtonian waveforms could
somewhat mimic some of the relativistic corrections, but a detailed
Fisher analysis of such effects is beyond the scope of this paper. 

The remainder of this paper is divided as follows:
Section~\ref{dynamics} deals with the dynamics of EMRBs in the
semi-relativistic approximation and justifies the use of this
approximation for these systems; Section~\ref{waveforms} reviews the
inclusion of higher-order multipolar corrections to the waveform
generation formalism; Section~\ref{numsim} describes the numerical
implementation of the equations of motion and the initial data used;
Section~\ref{comps} compares the orbital trajectories and waveforms;
Section~\ref{conclusions} concludes and points to future research.

In this paper, we denote the MBH mass by $\Mbh$ and its {\em
gravitational radius} by $\Rbh = 2G\Mbh/c^2$, where $c$ is the speed
of light and $G$ the Newtonian gravitational constant.  To simplify
some expressions we normalize masses with respect to $\Mmw = 4\times
10^6\,\Msun$, which is of the same magnitude as the mass of the MBH at
the center of the Milky Way \citep{Ghez:2005}.  The gravitational
radius can then be written as:
\begin{eqnarray}
  \Rbh &=& (3.82\times 10^{-7}~\textrm{pc}) \frac{\Mbh}{\Mmw} \,.
\end{eqnarray}
%


\section{EMRB Dynamics} \label{dynamics}

In this section, we discuss the description of the orbital motion.
Newtonian dynamics usually provides an adequate description of many
astrophysical sources of gravitational waves, at least from a
qualitative point of view.  However, for certain gravitational wave
sources, such a description is insufficient and relativistic effects
have to be considered.  For EMRB sources with pericenter distances
$r_p > 4\Rbh$ and velocities $v_{p}/c < 0.5$, the semi-relativistic
approximation to the equations of motion, in combination with a
multipolar description of the gravitational radiation, can adequately
model the dynamics and gravitational radiation, as we argue below.

The semi-relativistic approximation treats the motion of the SCO in
the point-particle limit as a geodesic of the MBH geometry, which is
justified based on the small mass-ratios associated with these
systems.  In this work, we adopt Cartesian Kerr-Schild coordinates,
$\{t,x^i\}$ ($i=1,2,3$), in which the MBH geometry is
time-independent, reflecting its stationary character, and tends to a
flat-space geometry in Cartesian coordinates far from the MBH.  We
denote the geodesic trajectory by $z^i(t)$, its spatial velocity by
$v^i(t) = dz^i/dt$, and its spatial acceleration by $a^i(t) =
dv^i/dt$.  The latter, in such a coordinate system, and by virtue of
the geodesic equations of motion, has the following form
\citep[e.g. see][]{Marck:1995kd}:
\begin{equation} \label{geodesics}
  a^i = F^i[v^j;\met^{}_{\mu\nu},\partial^{}_j\met^{}_{\mu\nu}] \,,
\end{equation}
where $\met^{}_{\mu\nu}$ ($\mu,\nu=0,1,2,3$) are the spacetime
components of the MBH metric. These equations describe the influence
of the spacetime curvature produced by the MBH and approach the
Newtonian equations of motion in the regime where $v/c = |v^i|/c \ll
1$ and $G\Mbh/(c^2\,r) \ll 1$ ($r=|x^i|$).

The effects from the self-gravity of the SCO can be neglected.  To see
this, consider the (Keplerian) orbital timescale, $T^{}_{\orb}$, in
comparison to a characteristic radiation-reaction timescale,
$T^{}_{\rr}$.  For the latter, we can use the timescale associated
with the rate of change of the semi-latus rectum, $p$, related to the
pericenter distance by $r_p = p/(1 + e)$, namely $T^{}_{\rr} \sim
p/|dp/dt|$.  The radiation-reaction timescales of the other orbital
elements are comparable or larger \citep[see,
e.g.,][]{Glampedakis:2005hs}. The ratio of these timescales is
\begin{equation} \label{eq:tscales}
  \frac{T^{}_{\orb}}{T^{}_{\rr}} \sim 2 \pi \mu
  \left(\frac{\Rbh}{2p}\right)^{5/2}\,,
\end{equation}
where $\mu = m/\Mbh$ is the mass ratio of the system and $m$ the SCO
mass.  It is evident that the radiation-reaction timescale is much
greater than the orbital timescale due to the extreme mass ratio, $\mu
\ll 1$, and because EMRBs have $p \gg \mu^{5/2} \Rbh$. In the unlikely
case that more accuracy is required, one could improve the analysis
through the use of ``Kludge'' waveforms \citep{Babak:2006uv}, which
have been shown to reproduce numerical results in the adiabatic
approximation accurately for EMRIs.

Formally, the orbital timescale used is not really the exact timescale
of orbital motion. This is because the mass distribution of a
MBH-embedded galaxy possesses a non-Keplerian potential that leads to
non-Keplerian orbits.  However, most EMRBs (by rate) have apocenters
that do not extend far into the stellar population, implying that the
contribution from the galaxy potential is minimal.  The orbits we
study in later sections have a contamination from the galactic
potential that is less than $2\%$ of the MBH mass.  Moreover,
\citet{Hopman:2006fc} rightly argue that the inner region is
statistically empty of stars, which is due to finite effects realized
at the small scales observed near the MBH.

Certain constraints may be derived on the size of $p$ and $r_{p}$ for
EMRB events. The most important constraint is derived from the
definition of EMRBs: orbits with sufficiently large orbital period
$T_{\orb} > T_{\cut}$.  Assuming a Keplerian orbit (which is a rough
assumption), this constraint translates to pericenter distances as
follows
\begin{equation}
  r_p > (7.98 \times 10^{-7}~\textrm{pc}) \frac{(1 -e)}{0.1}
  \left(\frac{\Mbh}{\Mmw}\right)^{1/3}
  \left(\frac{T}{T_{\cut}}\right)^{2/3} \,, \label{orbitalconstraint}
\end{equation}
where we have rescaled quantities assuming a typical eccentricity of
$e = 0.9$ \citep{Rubbo:2006dv,Rubbo:2007} and a typical MBH mass of
$\Mbh = \Mmw$.  In terms of geometrized units, such a constraint
translates roughly to $r_{p} > 2\Rbh$.

This constraint can be compared with the requirement that the SCO does
not get captured. Any object that enters the black hole event horizon
is captured, where the horizon is located (in Boyer-Lindquist
coordinates) at
\begin{equation}
  r_{\rm cap} = (1.91 \times 10^{-7}~\textrm{pc})\,
  \frac{\Mbh}{\Mmw}\, \left[ 1 + \sqrt{1 -
  \frac{\abh^2}{\Mbh^2}}\;\right]\,,
\end{equation}
where $\abh$ is the (Kerr) MBH spin parameter, related to its
intrinsic angular momentum by $\Sbh = G\Mbh\,\abh/c$, and bounded by
$\abh/\Mbh \leq 1$.  Thus, for a maximally spinning Kerr MBH ($\abh =
\Mbh$), $r_{\rm cap} = 0.5\Rbh$, while for a Schwarzschild
(non-spinning) MBH it is just $\Rbh$.  This condition tell us simply
that $r_p > r_{\rm cap}$, which is a condition superseded by the
constraint on the orbital period given in
equation~\eqref{orbitalconstraint}.  One could explore other possible
constraints \citep{Rubbo:2006dv, Rubbo:2007} but they are in general
superseded by equation~\eqref{orbitalconstraint}.

These constraints clearly exclude the ergosphere of the MBH ($r_{\rm
cap} < r \lesssim \Rbh$) where frame dragging effects are most
pronounced.  However, for EMRIs it has been argued
\citep{Glampedakis:2005hs} that orbits with $r_p < 10 \Rbh$ cannot be
considered Keplerian anymore, mainly due to precessional effects. This
statement can be made more quantitative by looking at the ratio of
first-order post-Newtonian (1~PN) predictions \citep{Blanchet:2002av}
to Newtonian ones (0~PN). For example, for the energy of a circular
orbit, this ratio scales as $7 \Rbh/(8 r_p)$, while for the perihelion
precession angle, the ratio scales as $3 \Rbh/(2 r_p)$, for
extreme-mass ratios.  Therefore, for orbits with pericenter passage
$r_p \sim 5\Rbh$, the 1~PN correction to the energy and the perihelion
precession angle is approximately $20 \%$ and $30 \%$ respectively,
relative to the Newtonian value. This indicates that, even for orbits
outside the ergosphere, relativistic effects are not necessarily
negligible.

The relativistic geodesic equations of motion introduce corrections to
the Newtonian motion that can be interpreted in terms of a black hole
\textit{effective} potential.  By comparing the Newtonian and
relativistic potentials one can see that the relativistic corrections
dominate over the centrifugal barrier at small distances from the
black hole center.  In this work we show that these relativistic
corrections can be sampled by EMRBs and hence, one should model these
systems accordingly.  Nevertheless, as we argued above, the
relativistic treatment of EMRBs does not need to be as sophisticated
as in the case of EMRIs, since radiation-reaction can be neglected.


\section{EMRB Waveforms} \label{waveforms}

In this section we describe how we extract gravitational waveforms
once we have integrated the geodesic equations of motion. We use a
multipole-moment wave generation formalism for slowly-moving objects
with arbitrarily strong internal gravity \citep{Thorne:1980rm,
Flanagan:2005yc, Glampedakis:2005hs}. In quasi-Newtonian and
semi-relativistic treatments, the radiation is modeled by the lowest
non-vanishing multipole moment: the mass-quadrupole.  To that order,
and for the case of a point-like object orbiting a MBH at a fixed
coordinate location, the plus and cross polarizations are given by
\citep{Misner:1973cw, Thorne:1980rm}
\begin{equation} \label{waves-quad}
  h_{+,\times} = \frac{2 G m}{r c^4} \; \epsilon^{ij}_{+,\times}
  \left( a_i z_j + v_i v_j \right),
\end{equation}
where $r$ is the (flat-space) distance to the observer and
$\epsilon^{ij}_{+,\times}$ are polarization tensors.  This expression
assumes, based on the slow motion approximation, that the change in
the acceleration with respect to time, the \textit{jerk}, $j^i =
da^i/dt$, is a small quantity.  More precisely, we are neglecting
terms of order $(v/c)^3$, or in other words, since the (quadrupole)
leading order terms are of order $(v/c)^2$, this implies a relative
error of order $v/c$.  

One can improve on this description for the gravitational radiation by
accounting for higher-order multipole moments. In this paper, we
consider the mass-octopole and current-quadrupole multipoles, which
require the knowledge of one more time derivative of the trajectory,
the jerk.  Adding these contributions to equation~\eqref{waves-quad},
the gravitational waveforms are given by~\citep{Thorne:1980rm}
\begin{eqnarray} \label{waves-oct}
  h_{+,\times} &=& \frac{2 G m}{r c^4} \; \epsilon^{ij}_{+,\times} \bigg\{
  a_i z_j + v_i v_j \nonumber\\
  &+& \frac{1}{c} \Big[ \left({\bf{n}} \cdot {\bf{z}}\right) \left(z_i j_j + 3 a_i v_j
  \right) + \left({\bf{n}} \cdot {\bf{v}}\right)
  \left( a_i z_j + v_i v_j \right) \nonumber\\ 
  &-& \left({\bf{n}} \cdot {\bf{a}}\right) v_i z_j - \frac{1}{2}
  \left({\bf{n}} \cdot {\bf{j}}\right) z_i z_j \Big] \bigg\}, 
\end{eqnarray}
where $n^i = x^i/r$ is a unit vector that points to the observer and
the vector product is the flat-space scalar product.  In this case, we
are neglecting terms of order $(v/c)^4$ and hence we are making a
relative error of order $(v/c)^2$ with respect to the leading order
quadrupole term.

The waveforms of equation~\eqref{waves-oct} are a truncated multipole
expansion, where we are neglecting the current-octopole,
mass-hexadecapole, and higher multipole moments. This expansion is
based on a slow-motion approximation which is valid for orbits whose
pericenter velocity is small relative to the speed of light.  For
closed circular orbits, we can use the Virial theorem to argue that
this is equivalent to requiring $r_{p} > \Mbh$. For a relativistic
EMRB event with $v_{p}/c = 0.4$, the maximum relative contribution of
the octopole to the quadrupole is of the order of~$40\%$, since the
octopolar term is of order $v/c$ smaller than the quadrupolar one. In
this paper, we shall study EMRBs from the sample of Milky Way sources
studied in~\cite{Rubbo:2006dv}. These sources have initial pericenter
distances of $r_{p} > 8 \Mbh$, thus justifying the use of a
low-multipolar expansion in the wave-generation formalism and the
neglect of radiation reaction effects in the orbital motion.


\section{Numerical Simulations} \label{numsim}

In this section we describe the EMRB simulations that were carried
out, including the choice of initial conditions.  The simulations
involve integrating the equations for geodesic motion around a Kerr
black hole, equation~\eqref{geodesics}, forward in time. \citep[For a
detailed exposition of Kerr geodesics see][]{Chandrasekhar:1992bo}.
Since Cartesian Kerr-Schild coordinates are used, the initial
conditions can be denoted by $(z^i_0,v^i_0)$.  The numerical
implementation does not use the Kerr geodesic constants of motion
(energy, angular momentum, and Carter constant) in order to reduce the
number of variables of the resulting system of ordinary differential
equations.  Instead, we have used the constants of motion to monitor
the accuracy of the time integration.  The integration accuracy is set
so that we obtain fractional errors for the constants of motion
smaller than one part in $10^{10}$.  The code uses a Bulirsh-Stoer
extrapolation method as the evolution algorithm \citep[see,
e.g.][]{Press:1992nr,Stoer:1993sb}.  We have also introduced in the
code the possibility of switching between Kerr geodesics and Newtonian
equations of motion.  The gravitational waveforms are then obtained
directly by applying expressions~\eqref{waves-quad} and
\eqref{waves-oct} to the numerically obtained trajectory $z^i(t)$.

Comparisons are carried out by choosing a representative relativistic
orbit within the allowed phase space for EMRB's.  We made the
following choices for the test case:
\begin{itemize}
\item The central MBH mass is $\Mbh = \Mmw$ and the SCO mass is $\Mco
  = 1~\Msun$, such that the mass ratio is $\mu = \Mco/\Mbh = 2.5
  \times 10^{-7} \ll 1$.

\item The MBH spin parameter is either $\abh=0$ (Schwarzschild) or $\abh
  = 0.998~\Mbh$ (Kerr).  The angular momentum is aligned along the
  $z$-axis and equal to either $S^z = 0$ or $S^z = 0.998 G\Mbh^2/c$.
  
\item The observer is located at $r_{obs} = 8$ kpc along the z-axis,
  which corresponds to the approximate distance from Earth to the center
  of the Milky Way \citep{Eisenhauer:2005}.
\end{itemize}
Furthermore, we make the following choices for the orbital initial
conditions: 
\begin{eqnarray} \label{orbit:ID}
  z^i_0 &=& \left(-1.59, 1.05, -0.185\right) \times 10^{-5}~{\textrm{pc}}, 
  \nonumber\\
  v^i_0 &=& \left(1.70,-2.89,0.510\right)\times 10^{4}~{\textrm{km
  s}}^{-1}\,.
\label{testorbit}
\end{eqnarray}
The initial conditions are such that $r_{0} = |z^i_0| = 50 \Rbh = 1.91
\times 10^{-5}$~pc, and $|v^i_0| = 0.11\,c = 3.39 \times
10^{4}~{\textrm{km s}}^{-1}$.  The orbital plane is inclined by
$10^{\circ}$ with respect to the $x-y$ plane to demonstrate the
effects of orbital plane precession, which only occurs for spinning
MBHs. Since this paper is concerned with the effect of relativistic
corrections to EMRB events, we choose to give all orbits the same
initial conditions. The possibility of relaxing this choice and its
effect on the conclusions derived in this paper shall be discussed in
a later section.

Although the test orbit is in the phase space of EMRB events studied
in~\citet{Rubbo:2006dv}, one might worry that it is too relativistic
to actually have a significant probability to occur in nature.  In 
particular, one can think that the SCO may be tidally disrupted. To
address this question, let us consider a Newtonian description of the 
central potential, which leads to the following values of the pericenter 
distance and velocity:
\begin{eqnarray}
  r_{p} &=& 6.45~\Rbh = 2.48 \times 10^{-6}~{\textrm{pc}},
  \nonumber \\
  |v_{p}| &=& 0.384\,c = 1.15 \times 10^{5}~{\textrm{km s}}^{-1} \,.
\end{eqnarray}
One might worry that stars might be tidally disrupted with such small
pericenters. However, as shown by \citet{Hopman:2006fc}, most SCOs in
EMRB scenarios consist of stellar-mass black holes, which cannot be
tidally disrupted.

Such relativistic orbits are actually naturally occurring in the phase
space of possible EMRBs studied in~\citet{Rubbo:2006dv}. Of all
orbits in the allowed EMRB phase space considered
in~\citet{Rubbo:2006dv, Rubbo:2007}, $6\%$ is contained within a small
$6$-dimensional phase space volume centered on the test
orbit~\footnote{In other words, $6\%$ of the test orbits considered to
  be EMRBs in ~\citet{Rubbo:2006dv} are close in phase space to our
  test orbit. This does not imply that the probability of such a test
  orbit actually occurring in nature is $6 \%$, since any element of
  phase space may have small overall probabilities, even down to
  $10^{-7} \%$.}. Furthermore, the test case shown possesses a short
orbital timescale, bursting approximately $150$ times per year. Such
events with small orbital timescale were shown to dominate the EMRB
event rate in~\citet{Rubbo:2006dv}. It is in this sense that the test
orbit studied here is {\emph{typical}} or representative of EMRBs.

\begin{figure}
\includegraphics[scale=0.375,clip=true]{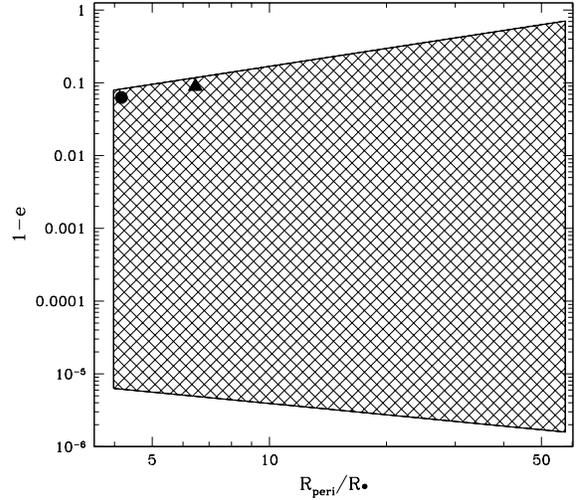}
\caption{\label{ID} Plot of the set of possible EMRB orbits as
  computed in \citet{Rubbo:2006dv} in the pericenter
  distance-eccentricity plane. The pericenter distance is given in
  units of the gravitational radius $\Rbh$. The initial conditions
  for the test [Eq.~(\ref{testorbit})] and the extreme 
  [Eq.~(\ref{extremeorbit})] orbits are indicated by a triangle and
  a circle respectively.}
\end{figure}

The relative location of the test orbit in the pericenter
distance-eccentricity plane of the phase space of allowed
EMRBs~\citep{Rubbo:2006dv} is presented in Figure~\ref{ID} (triangle.)
The eccentricity was here calculated assuming a Newtonian orbit and
the pericenter separation is given in gravitational radii, $\Rbh$.
Although the test orbit has a large eccentricity, its apocenter is
small enough ($r_{a} \lesssim 150 \Rbh \approx 6 \times 10^{-5}$~pc)
that the contribution from the surrounding stellar population to the
potential can be neglected.  In general, the left side of the figure
corresponds to highly relativistic orbits with large pericenter
velocities and small pericenter distances.  Orbits with pericenter
velocities $|v_p| > 0.25\,c = 3 \times 10^4~{\textrm{km s}}^{-1} $
account for approximately 50\% of the possible orbits within the phase
space studied in~\citet{Rubbo:2006dv}.

Although the test orbit is a good source to demonstrate the
differences between the Newtonian and relativistic treatments, we
could have chosen an even more relativistic one. Such an event would
still be classified as an EMRB in a Newtonian treatment, but it would
border with the definition of a continuous source.  An example of such
an extreme orbit is shown with a circle in Figure~\ref{ID}, to the
left of the test orbit (triangle).  This extreme orbit possesses the
following initial conditions:
\begin{eqnarray}
  z^i_0 &=& \left(-1.81,0.6,-1.06\right) \times 10^{-6}~{\textrm{pc}},
  \nonumber \\
  v^i_0 &=& \left(1.72,-1.78,0.31\right) \times 10^{5}~{\textrm{km
  s}}^{-1},
  \label{extremeorbit}
\end{eqnarray}
where $r_p = 4\,\Rbh = 1.53 \times 10^{-6}$~pc and $|v_p| = 0.49c =
1.46 \times 10^5~{\textrm{km s}}^{-1}$ for a Newtonian potential. We
will study such an extreme orbit at the end of the next section as an
example of a limiting relativistic case.


\section{Comparison of Trajectories and Waveforms} \label{comps}

In this section we compare the results obtained for both the orbital
motion and the gravitational radiation emitted by an EMRB event using
both the Newtonian and relativistic description.  Since the {\em plus}
and {\em cross} polarization waveforms present similar features, we
only plot the {\em plus} polarization waveforms.  In the remainder of
this section we use the following nomenclature: a quadrupolar
(octopolar) Newtonian waveform is one that was calculated using the
quadrupole (octopole) formula and Newtonian equations of motion; a
quadrupole (octopole) Schwarzschild waveform is one that was
calculated using the quadrupole (octopole) formula and the geodesic
equations of motion with no spin ($\abh=0$); a quadrupole (octopole)
Kerr waveform is one that was calculated using the quadrupole
(octopole) formula and the geodesic equations of motion with spin
$\abh = 0.998 \Mbh$.

\subsection{Orbital Trajectories}

\begin{center}
\begin{figure}
 \includegraphics[scale=0.85,clip=true]{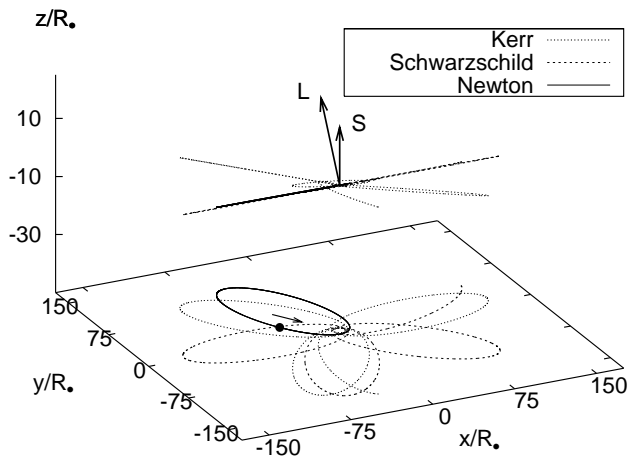}
\caption{\label{Orbits} Trajectories for the SCO, with initial
  conditions given by~\eqref{orbit:ID}, corresponding to a Newtonian
  description (solid line) and relativistic descriptions with no spin
  (dashed line) and with spin (dotted line).  The MBH is located at
  the origin and the vectors $L$ and $S$ denote the initial orbital
  angular momentum and the MBH spin respectively. }
\end{figure}
\end{center}

Let us begin by comparing the trajectories obtained in our
simulations.  In Figure~\ref{Orbits}, we plot the test orbit
corresponding to a Newtonian treatment (solid line) and the one
corresponding to a relativistic treatment without spin (dashed line)
and with spin (dotted line).  The dot and arrow indicate the initial
location and velocity projected onto the $x$-$y$ plane. The MBH is
located at the origin of the coordinates, and the vectors denoted by
$L$ and $S$ describe the direction of the initial orbital angular
momentum and the MBH spin respectively.  In the relativistic
description there are precessional effects in the SCO orbit that can
be clearly observed in Figure~\ref{Orbits}.  These precessional
effects are: pericenter precession about the orbital angular momentum
axis, which acts in the initial orbital plane; and frame-dragging
precession about the total angular momentum axis, which acts out of
the initial orbital plane.  While the former always occurs in a
relativistic treatment, the latter is present only in the spinning
case.

Different relativistic precessional effects are generally of different
magnitude. These effects are usually inversely proportional to the
pericenter distance, or equivalently proportional to the pericenter
velocity of the SCO.  Precession out of the initial orbital plane,
however, is smaller than precession in the orbital plane by a relative
factor of order $v_{p}/c$ and it is directly proportional to the spin
of the MBH. In terms of post-Newtonian theory \citep[see,
e.g.][]{Blanchet:2002av}, the pericenter advance is described by
1st-order post-Newtonian corrections to the equations of motion (order
$(v/c)^2$ relative to the Newtonian acceleration), while precession
off the orbital plane is due to spin-orbit and spin-spin couplings
that correspond to $1.5$ and $2$-order corrections (order $(v/c)^3$
and $(v/c)^4$ relative to the Newtonian acceleration.)  Therefore,
since EMRBs are characterized as events with small to moderate
pericenter velocities, precession out of the initial orbital plane is
small to moderate relative to pericenter advance, even for maximally
spinning MBHs.

We can estimate the precession rate by comparing the Newtonian and
relativistic trajectories.  For the test orbit considered, we find
that the rate in the orbital plane is roughly $\pi/3$ radians per
orbit for the non-spinning case and $2\pi/3$ radians per orbit for the
spinning one. These precessional effects have been studied extensively
in the context of EMRIs \citep[see, e.g.][]{Schmidt:2002} and also
specifically for S-stars in the central region of our Galaxy in
\citet{Kraniotis:2007}.  Nonetheless, these effects have not been
previously analyzed in the context of EMRBs, since previous studies
employed a quasi-Newtonian treatment. 

\subsection{Waveforms}
\begin{figure}[tl]
  \includegraphics[scale=0.35,clip=true]{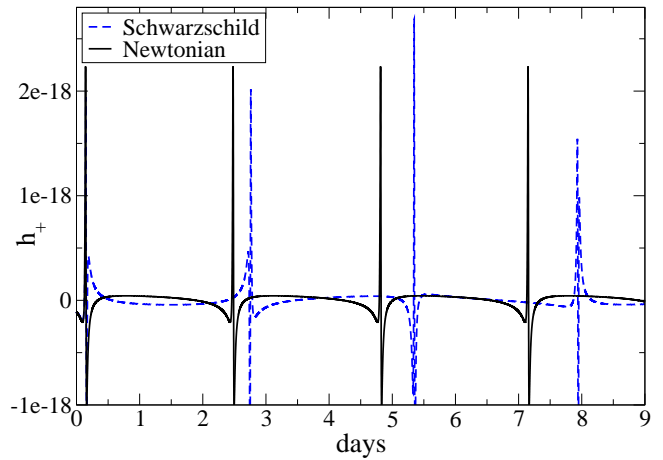}
\caption{\label{hp1} EMRB waveforms ({\em plus} polarization):
The Newtonian waveform corresponds  to the solid line and
the Schwarzschild one to the dashed line.}
\end{figure}

\begin{figure}[tr]
\includegraphics[scale=0.35,clip=true]{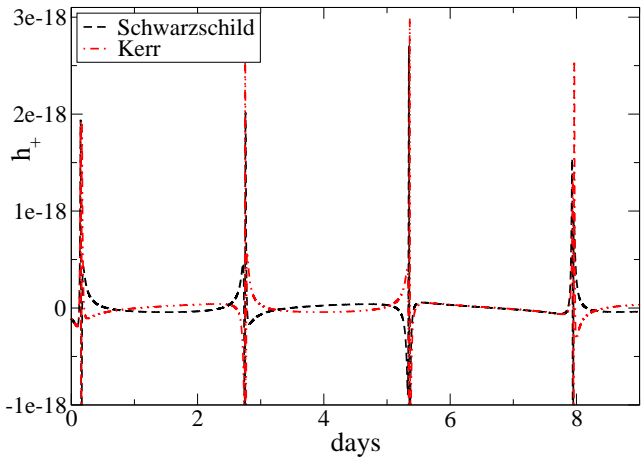}
\caption{\label{hp2} EMRB waveforms ({\em plus} polarization):
The Schwarzschild waveform corresponds to the dashed
line and the Kerr one to the dot-dot-dashed line.
The dephasing of the waveforms can be best observed during the
silent transitions between bursts. For example, in the first silent
transition the waveforms are roughly $\pi$ radians out of phase, while
in the third one they are in phase.}
\end{figure}

Let us now analyze how the differences in the SCO trajectories
translate into different signatures in the waveforms.  In
Figure~\ref{hp1} we plot the quadrupole Newtonian and Schwarzschild
waveforms (solid and dashed lines respectively), while in
Figure~\ref{hp2} we plot the quadrupole Schwarzschild and Kerr
waveforms (dashed and dot-dot-dashed lines respectively.) There are
three main differences between the Newtonian and the relativistic
waveforms: a modulated phase change, an amplitude change, and a time
of arrival change.  The changes in amplitude and time of arrival are
due to the test particle experiencing a larger ``force'' of attraction
as it approaches the black hole.  Quantitatively, this increase in
force is due to the presence of ($r_p^{-n}$)-contributions to the
relativistic corrections to the central potential (with $n$ a real
positive number.)


Gravitational wave interferometers are most sensitive to the phase,
which is clearly different for Newtonian and relativistic waveforms.
The dephasing present in Figures~\ref{hp1} and \ref{hp2} parallels the
orbital dephasing discussed earlier (the gravitational-wave and
orbital frequencies are intimately related) and leads to an amplitude
modulation. In terms of the gravitational wave phase, both
Figures~\ref{hp1} and \ref{hp2} show a dephasing of $\pi/6$ radians
per cycle.  This can be seen after the third burst where the waveforms
are back in phase. In fact, there is a significant dephasing even
between the relativistic waveforms due to the effect of the MBH spin.
If a cursory examination by eye can detect the difference in the
waveforms due to differences in the nature of the central potential,
it stands to reason that strong-field EMRB waveforms might allow us to
probe of the spacetime near a MBH. We should note, however, that no
work has yet been done to find best-fit parameters for Newtonian
waveforms that maximize the correlation with relativistic ones. In
other words, it might be possible to mimic some of the relativistic
corrections by choosing different initial data for the Newtonian
waveforms, but such a study is beyond the scope of this paper.

\begin{figure}[tl]
\includegraphics[scale=0.33,clip=true]{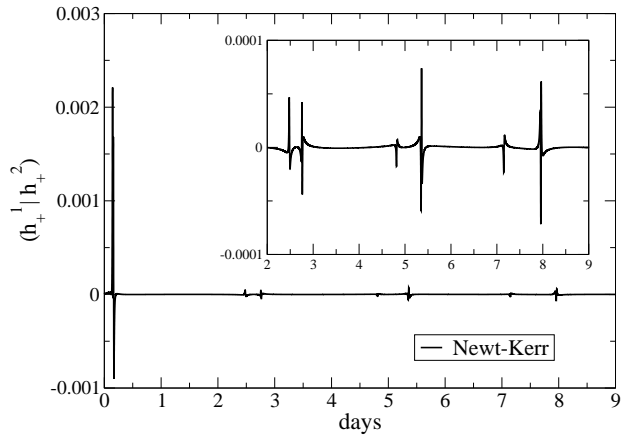}
\caption{\label{power1} Plot of the overlap integrand of
  equation~\eqref{corr} with $h_{1}$ given by the quadrupole Kerr
  waveform and $h_{2}$ by the quadrupole Newtonian one.  The inset
  zooms to a region near the small peaks.}
\end{figure}

\begin{figure}[tr]
\includegraphics[scale=0.33,clip=true]{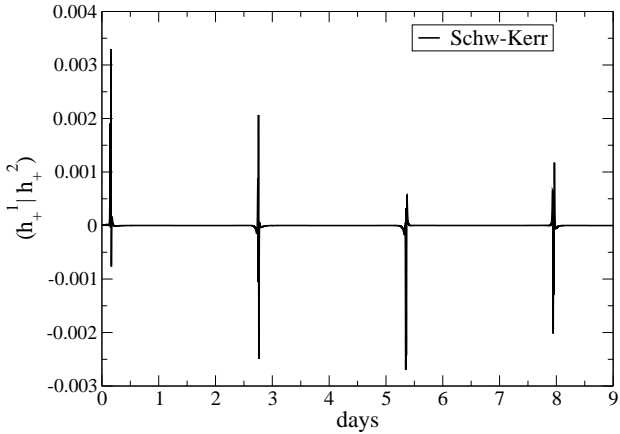}
\caption{\label{power2} Plot of the overlap integrand of
  equation~\eqref{corr} with $h_{1}$ given by the quadrupole Kerr
  waveform and $h_{2}$ by the quadrupole Schwarzschild one.}
\end{figure}

The difference in dephasing can be better studied by calculating the
signal overlap,
\begin{equation} \label{corr}
  (h_{1} | h_{2}) = \frac{\int_{0}^{T} h_{1}(t)
  h_{2}(t) dt}{\sqrt{ \int_{0}^{T} h_{1}^{2}(t) dt} \;
  \sqrt{\int_{0}^{T} h_{2}^{2}(t) dt}} \;.
\end{equation}
The overlap indicates how well a signal $h_{1}$ can be extracted via
matched filtering with a template $h_{2}$~\footnote{The overlap is
  given here in the time domain, but an analogous representation in
  the frequency domain could also be used. Such an expression in the
  frequency domain can be derived through Parseval's theorem.}.  In
Figures~\ref{power1} and~\ref{power2} we plot the normalized integrand
as a function of time, with $h_{1}$ given by the quadrupole Kerr
waveform and $h_{2}$ given by either the quadrupole Newtonian or
Schwarzschild waveforms.  Observe that neither the Newtonian nor the
Schwarzschild waveforms match well with the Kerr waveform.  Moreover,
note that the correlation with the Newtonian waveform deteriorates
greatly after only the first cycle. The integral of
equation~\eqref{corr} gives the correlation between different
waveforms over nine days (four bursts): for the Newtonian and Kerr
plus-polarized waveforms it is $9.6 \%$; for the Schwarzschild and
Kerr plus-polarized waveforms it is $-6.3 \%$.  As a point of
comparison, a substantial signal overlap should be~$\gtrsim 95 \%$. As
already mentioned, the same initial conditions were chosen for both
the Newtonian and relativistic orbits, such that their waveforms would
be both initially in-phase and any dephasing due to relativistic
effects could be clearly studied. However, such a choice forces the
SCO to pass through periapsis at slightly different times, because in
the relativistic case this object experiences a ``deeper'' potential.
Such a difference in timing degrades the overlap somewhat and could,
in principle, be ameliorated by choosing different initial conditions
for the Newtonian evolution, {\textit{i.~e.~}} by maximizing the
overlap over all orbital parameters, but such a study is beyond the
scope of this paper.

Figures~\ref{power1} and~\ref{power2} provide some evidence that the
use of a relativistic waveform might be required for the data analysis
problem of extracting EMRB signals. Such expectations are somewhat
confirmed in Table~\ref{table_orbs}, where we present the correlation
between Newtonian and Kerr plus-polarized waveforms integrated over
both a single day (one burst) and nine days of data (several bursts)
for a sample of different EMRB orbits~\footnote{Here we use the
  frequency representation of the correlation calculation, employing
  the Fourier transform of the waveforms.}. Both a single and several
bursts should be studied because, in principle, parameter adjustments
might mitigate the between-burst dephasing, but probably not the
in-burst dephasing. The orbits chosen were taken directly from the
allowed EMRB phase space of~\citet{Rubbo:2006dv} and possess different
initial positions and velocities, leading to different eccentricities,
initial inclination and orbital periods. The orbital period can be
used to classify the orbits into highly-bursting (burst more than once
per week) and slowly-bursting (burst less than once per week.)  For
highly-bursting EMRBs, the average correlation is $\sim 27 \%$ when
all nine days of data (several bursts) are used, while it is $\sim 93
\%$ when only one day of data (single burst) is used. For the
slowly-bursting EMRBs we considered here, there is actually only one
burst per week and its correlation is $\sim 85 \%$. These results
indicate that accumulating precession effects lead to a significant
loss of correlation between Newtonian and Kerr waveforms if bursts are
to be connected. Moreover, we see that even for a single burst the
shape of the relativistic waveforms is sufficiently different from its
Newtonian counterpart to lead to a significant mismatch (in-phase
dephasing). If one maximizes the correlation over intrinsic orbital
parameters it might be possible to increase the correlation somewhat,
but again that is to be studied elsewhere.


\begin{figure}
\includegraphics[scale=0.33,clip=true]{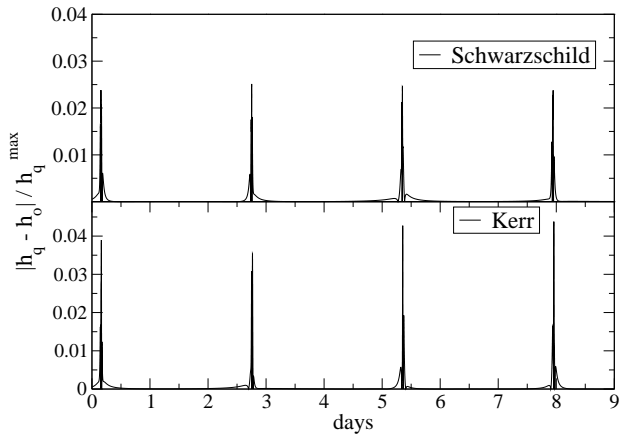}
\caption{\label{diff} Plot of the absolute magnitude of the difference
  between the quadrupole and octopole Schwarzschild (upper panel) and
  Kerr (lower panel) waveforms. The difference is normalized to the
  maximum value of the first peak of the quadrupole waveform.}
\end{figure}

Let us now focus on the differences in the waveforms when they are
calculated with the quadrupolar approximation versus the
quadrupolar-octopolar one. In Figure~\ref{diff} we plot the absolute
value of the difference between the octopole and quadrupole waveforms
as a function of time for a Schwarzschild (upper panel) and a Kerr
(lower panel) central potential. The difference is normalized to the
maximum of the first peak of the quadrupole waveform, since other
peaks have approximately the same maximum.
Observe that the inclusion of higher-order multipoles does not affect
the phasing of the waveforms, but only the amplitude, which is in
general different by roughly 4\% relative to the quadrupole waveform
for the Kerr test case. At first sight, this result is in disagreement
with the expectation that the octopolar correction is at most
$\lesssim 40 \%$ of the quadrupolar one. Note, however, that the $40
\%$ estimate is an order-of-magnitude {\emph{upper limit}}, since the
octopole correction is dependent on the location of the observer
relative to the trajectory, velocity, acceleration and jerk vectors.
For the test case, where the observer is located on the $\hat{z}$-axis
and the orbit is initially inclined by $10^{\circ}$, the octopolar
change is reduced by approximately an order of magnitude, since
initially $(\mb{n} \cdot \mb{z}) \approx (\mb{n} \cdot \mb{v}) \approx
(\mb{n} \cdot \mb{a}) \approx (\mb{n} \cdot \mb{j}) \approx 0.1$. In
Table~\ref{table_obs}, we present the approximate maximum difference
between octopolar and quadrupolar waveforms as a function of observer
location, focusing only on the first burst of radiation. The location
of the observer is rotated about the $\hat{y}$-axis on the
$\hat{x}-{\hat{z}}$ plane, always at a fixed radial distance of 8~kpc.
Note that for some observers the difference is larger and, in fact, of
the order of 40\%, since the dot products are closer to unity.  These
results are thus consistent with the expectation that the $n$-th
multipolar contribution cannot in general be larger than order
$(v/c)^n$ relative to the quadrupolar leading term.

\subsection{Data Analysis}

In order to quantify some of our statements about the change in phase
and amplitude, we calculated the SNR for the relativistic waveforms
via the standard formula
\begin{equation}
  \rho^2 = 4 \int_{0}^{\infty} \frac{|\tilde{h}(f)|^{2}}{S_{n}(f)} \;
  df \;,
\end{equation}
where the tilde denotes the Fourier transform and $S_{n}(f)$ is the
one-sided power spectral noise density.  Here we employ the Online
Sensitivity Curve Generator \citep{Larson:2000} with the standard LISA
settings and the inclusion of the white-dwarf background contribution.
When calculating SNRs, we set the observation time to roughly nine
days, so as to include multiple bursts in our single SNR value.

The inclusion of relativistic corrections in the trajectories has a
dramatic impact in the SNR. We find that the Schwarzschild waveform
increases the SNR by a factor of approximately $59\%$, while the Kerr
waveform increases it by $162\%$, relative to the Newtonian SNR.  The
difference in SNR is because the relativistic orbits experience a
deeper effective potential and, thus, the interaction timescale is
smaller.  Therefore, the inverse of the interaction time, $f_{\star} =
v_p/r_p$, is larger for the Schwarzschild and Kerr waveforms relative
to the Newtonian one.  As a result, the Fourier power is shifted to
higher frequencies, where LISA is more sensitive.

The SNR behaves similarly for other EMRB orbits with different orbital
periods, eccentricities and pericenter parameters. This can be
observed in Table~\ref{table_orbs} where we present the SNR difference
between Newtonian and Kerr waveforms for different EMRB orbits for a
single and several bursts. Highly-bursting EMRBs lead to a large
change in the SNR over a week of data, since they experience the
depths of the effective potential several times. Per burst, the change
in SNR can range from one, to ten or even ninety percent, depending on
the inclination angle of the orbit, the pericenter distance and other
orbital parameters. Also note that the orbits presented in the table
are not as relativistic as the test case, which is why the SNR
difference is smaller. This study seems to indicate that the SNR is in
general somewhat larger for relativistic waveforms, specially when
several burst are taken into account.~\footnote{Naive intuition might
  suggest that the change in the SNR should scale like the square root
  of the number of bursts, but this is not necessarily correct. First,
  different orbits possess different beaming patterns on the sky due
  to precession. Second, the starting frequencies of these bursts is
  very close to the limit of LISA's sensitivity band [$10^{-5}$ Hz].
  As the orbits burst, precession somewhat increases the frequency of
  the waveform, forcing different bursts to contribute different
  amounts to the SNR.}  Consequently, the event rate calculated in
\citet{Rubbo:2006dv, Rubbo:2007}, which in particular summed over all
bursts in one year of data, is an underestimate for their galaxy
model, because some of the systems with a Newtonian SNR $\lesssim 5$
should have been added to the detectable event rate.  However, the
uncertainty in the event rate is still probably dominated by
astrophysical uncertainties and not by the dynamics modeling.

Conversely, the inclusion of higher multipole moments to the wave
generation formalism has little to no effect in the SNR. In the
previous section we showed that there was $\approx 4\%$ difference
between the octopole and quadrupole waveforms relative to the maximum
of the first peak of the quadrupole waveform. We further showed that
this difference depends on the location of the observer (see
table~\ref{table_obs}), but for the test case it does not exceed a
maximum of 40\%, which agrees with the multipole-ordering argument
previously described.  However, we also pointed out that the amplitude
difference is confined to sharp peaks in the time domain. Such a
confined change in the waveform amplitude leads to a Fourier power
being dispersed over a large frequency region, including outside the
LISA band.  As a result, there is a correspondingly small change in
the SNR: of the order of $\approx 1\%$ relative to the quadrupolar
formalism. Such a result is in agreement with the analysis
of~\cite{Babak:2006uv}, which was carried out for EMRIs. Therefore, we
see that the change in SNR is primarily dominated by the modifications
introduced in the geodesic description of the equations of motion, and
not in the octopolar correction to the waveform generation.

The analysis presented in this section, in particular
Figure~\ref{power1}, makes it clear that relativistic corrections to
the waveforms accumulate with multiple bursts. In other words, over a
single burst (pericenter passage), a quadrupolar waveform calculation
using Newtonian dynamics might be sufficient.  However, if one wants
to estimate parameters associated with the central MBH, then multiple
bursts might be necessary to associate the events to a single SCO
trajectory.  In terms of data analysis, for a detection search it is
simpler to look for a single burst using techniques such as excess
power and wavelet decompositions \citep[e.q.  see][]{Anderson:2000yy,
  Klimenko:2004, Stuver:2006, Camarda:2006}. For estimating MBH
parameters, the results of this paper suggest that multiple bursts
might have to be connected. For this to occur, a single template may
be used, but as our results indicate, the template will need to
incorporate the effects of general relativity.~\footnote{A note of
  caution should be added here, since a detailed study of the
  maximization of the overlap with respect to orbital parameters in
  the Newtonian waveform has not yet been carried out. Indeed, it
  might be possible to mimic some of the relativistic effects with
  Newtonian waveforms with varying parameters, but such mimicking is
  most probably not possible for highly relativistic EMRBs.}

At this junction, we should comment on some of the caveats in the
conclusions derived from the analysis presented here. First, in this
paper we have primarily concentrated on the question of characterizing
the gravitational waves through the study of the SNR and overlap. In
this study, however, we have not maximized these data analysis
measures with respect to (intrinsic or extrinsic) orbital parameters.
Although it might be possible to improve the SNR and overlap via
parameter maximization, our study suggests that the introduction of
relativistic effects, such as precession, lead to a clear imprint on
the waveform that might be difficult to mimic with a purely Newtonian
waveform irrespective of its parameters. Second, in this paper we have
only touched the iceberg of the signal characterization and parameter
estimation problem. A possible route to study this problem is through
a numerical Fisher analysis, with the complications derived from the
fact that the waveforms are known only numerically. Furthermore,
increasing the complexity of the waveform will also increase the
computational cost of these studies and, thus, it might be interesting
to investigate whether it is possible or advantageous to search for
individual bursts with similar frequency and identify them as
belonging to the same physical system.  These issues are beyond the
scope of this paper, but they should be addressed in future
investigations.

\begin{figure}
\includegraphics[scale=0.33,clip=true]{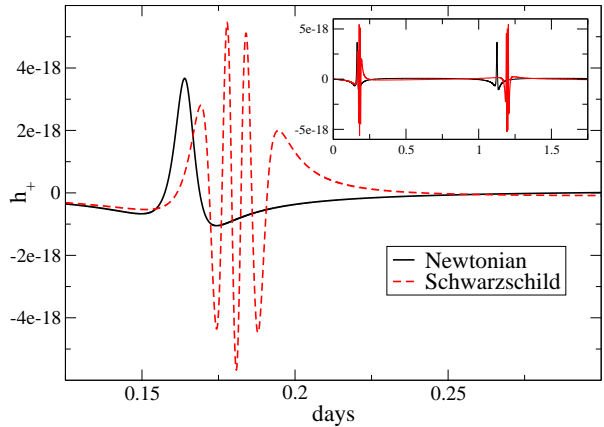}
\caption{\label{extreme} Plot of the quadrupole Newtonian (solid) and
  Schwarzschild (dashed) gravitational waveform as a function of
  time.}
\end{figure}

Setting these caveats momentarily aside, let us conclude with some
discussion of the extreme relativistic case introduced earlier. As we
have seen, relativistic corrections can introduce strong modifications
to EMRB waveforms, which depend on how relativistic the EMRB event is
and, in particular, on the pericenter velocity.  The corrections are
particularly strong for the class of EMRBs that inhabit the boundary
between EMRBs and EMRIs, defined by the $T_{\cut} = 3 \times 10^{4}$~s
value, corresponding to the period between apocenter passages.  An
example of such an event is the extreme orbit discussed in
Section~\ref{numsim}, whose waveform is shown in Figure~\ref{extreme}.
Observe that a simplistic Newtonian description misses the rich
structure, in which the SCO whirls twice about the black hole before
zooming out to apocenter again. This behavior is missed entirely when
we evolve the orbit with the Newtonian equations of motion, even
though the same initial conditions were used. Even though in the
previous cases a Newtonian waveform might be able to extract
relativistic events by adjusting intrinsic parameters, such is
definitely not the case for the highly-relativistic event of
Fig.~\ref{extreme}, since no choice of parameters in the Newtonian
waveform can reproduce its multiple-peak structure.

Due to their whirling behavior, the extreme orbit waveform resembles
the zoom-whirl events often mentioned in the EMRI literature
\citep{Hughes:2001:a, Hughes:2001:b, Glampedakis:2002}.  However, the
event is still an EMRB and not an EMRI because the period between
apocenter passages is too long.  For our galactic model, we find the
probability of a small region of phase space around this orbit to be
rather high, 10\%.  If this EMRB is detected with sufficiently high
SNR it seems plausible that a parameter estimation analysis would
allow for a determination of the background parameters, such as the
black hole spin. \citet{Barack:2004a} have already investigated LISA's
ability to measure MBH properties using approximate EMRI signals.
They found that, depending on the actual orbital parameters, it will
be possible to measure the MBH spin with fractional errors of
$10^{-3}$ to $10^{-5}$.  This high precision measurement is the result
of observing up to $\sim\!10^{6}$ complete orbits. Conversely, it is
very unlikely that EMRB measurements will be able to match the
measurement capabilities of EMRI signals, since only a few bursts will
probably be available. Whether accurate parameter extraction is
possible can only be determined with a more detailed data analysis
investigation of EMRBs.

\section{Conclusions} \label{conclusions}

We have studied the effects of relativistic corrections on the
gravitational waves produced by EMRBs. These events originate from
long period orbits of a SCO around a MBH, leading to large-amplitude,
quasi-periodic gravitational wave bursts. Using a more accurate
relativistic treatment of the phenomenon, we have improved on the
waveforms and trajectories relative to previous work.  The orbital
trajectories were corrected by accounting for the spacetime curvature
of the system for Schwarzschild and Kerr MBHs. The waveform generation
was corrected by accounting for the next order term in the multipolar
expansion of gravitational radiation.

We found that relativistic corrections change the waveform shape
relative to its Newtonian counterpart. One of the most significant
changes was found to be an amplitude-modulated dephasing, produced by
the relativistic corrections to the orbital trajectory and, in
particular, by relativistic precessional effects. Other effects
included a change in the amplitude of the waveform, partially produced
by the inclusion of higher-order terms in the gravitational wave
generation scheme.

The magnitude of the relativistic corrections was found to be directly
proportional to the pericenter velocity of the orbit, as expected.
Surprisingly, we estimated that at least $50 \%$ of the orbits
analyzed in \citet{Rubbo:2006dv,Rubbo:2007} acquire relativistic
velocities and, thus, non-negligible relativistic corrections. We
investigated these corrections in detail by choosing a test orbit,
that is representative of the kind that dominated the event rate
calculation of~\citet{Rubbo:2006dv}. We also studied a limiting case
of a highly relativistic EMRB and found that it whirls more than once
around the MBH before zooming back to apocenter and becoming silent
again.

We have also discussed the possible consequences that relativistic
effects might have on the detection and parameter estimation of
gravitational waves from EMRBs by LISA, namely a change in the SNR and
loss of overlap. These changes are mainly due to the relativistic
treatment of the equations of motion, while a quadrupolar wave
generation formalism seems to suffice. This finding is relevant
particularly to match filtering searches, where a Newtonian treatment
of the orbit might lead to a deterioration of confidence limits.
Furthermore, our study suggests that, given an EMRB gravitational wave
detection, it might be plausible to extract or bound the spin of the
central potential with a template that takes into account the Kerr
character of the MBH. Other astrophysical consequences include a
possible increase in the event rate, which implies that the rates of
\citet{Rubbo:2006dv,Rubbo:2007} and \citet{Hopman:2006fc} might be
lower limits, although these estimates are still dominated by
uncertainties in the astrophysical modeling for the host galaxy.

In addition to the astrophysical modeling, future research should
tackle the details of the EMRB data analysis and signal extraction
issues put forward above. Based on the results of this paper, one may
explore the possibility of testing alternative theories of gravity
with EMRBs by performing matched filtering with templates from
alternative theories \citep{will:1998:bmo, scharre:2002:tsg,
  will:2004:tat, berti:2005:tgr, berti:2005:esb}.  Another possible
avenue for future research is the study of confidence limits with
which the spin of the central MBH can be measured.

This research could then be used to examine whether EMRB events can
distinguish between a pure Kerr MBH and some other spacetime. Such
studies have already began with the analysis
of~\citet{Collins:2004ex},~\citet{Glampedakis:2005cf}
and~\citet{Barausse:2006vt}, where comparisons between a Kerr and
other non-Kerr spacetimes were performed. Such studies could be
extended to the perturbed Kerr solution found by~\citet{Yunes:2005ve},
where the perturbation is parameterized by the Weyl tensor of the
external universe and could represent some external accretion disk,
planetary system or some other compact object. Ultimately, these
investigations will decide whether EMRB events are worth studying in
further detail by future gravitational wave observatories.


\acknowledgments

The authors would like to thank Ben Owen for reading the manuscript
and providing useful comments, as well as the anonymous referee for
providing insightful suggestions. We would also like to acknowledge
the support of the Center for Gravitational Wave Physics funded by the
National Science Foundation under Cooperative Agreement PHY-01-14375,
and support from NSF grants PHY 05-55628, PHY 05-55436, PHY 02-18750,
PHY 02-44788, PHY 02-45649 and PHY 00-99559.  K.~H.~B.~and L.~R.~also
acknowledge the support of NASA NNG04GU99G, NASA NN G05GF71G.
C.~F.~S. acknowledges the support of the Natural Sciences and
Engineering Research Council of Canada.


\clearpage 
\begin{center}
\begin{deluxetable}{c c c c c c c c c}
  \tablewidth{0pt} \tablecaption{\label{table_orbs} SNR and overlap for
    different EMRB orbits.}  \tablehead{ \colhead{$r_p$ [$\mu$pc]} &
    \colhead{$v_p/c$} & \colhead{$P$ [yrs]} &
    \colhead{$e$} & \colhead{$\#$ Bursts} & \colhead{$\frac{|\Delta
    \rho|}{\rho}$} & \colhead{$\frac{|\Delta
    \rho^{(1)}|}{\rho^{(1)}}$} & \colhead{$(h_1|h_2)$} 
    & \colhead{$(h_1|h_2)^{(1)}$} }  
\startdata 
$7.8$ & $0.196$ & $0.0042$ & $0.603$ & $6$ & $0.38$ & $0.049$ & $0.27$ & $0.98$\\[1mm]
$7.5$ & $0.207$ & $0.0055$ & $0.684$ & $5$ & $0.13$ & $0.0007$ & $0.20$ & $0.88$\\[1mm]
$7.2$ & $0.217$ & $0.0074$ & $0.749$ & $4$ & $0.22$ & $0.057$ & $0.31$ & $0.90$\\[1mm]
$7.1$ & $0.220$ & $0.0086$ & $0.777$ & $3$ & $0.49$ & $0.061$ & $0.33$ & $0.99$\\[1mm]
      &                      &          &         &     &     \\[1mm]
$13$  & $0.163$ & $0.0450$ & $0.859$ & $1$ & $\cdot$ & $0.002$ & $\cdot$ & $0.90$\\[1mm]
$8.0$ & $0.217$ & $0.1953$ & $0.968$ & $1$ & $\cdot$ & $0.033$ & $\cdot$ & $0.97$\\[1mm]
$10$  & $0.193$ & $0.7683$ & $0.984$ & $1$ & $\cdot$ & $0.009$ & $\cdot$ & $0.72$\\[1mm]
$12$  & $0.173$ & $3.0407$ & $0.992$ & $1$ & $\cdot$ & $0.015$ & $\cdot$ & $0.80$\\[1mm]
\enddata \tablecomments{In this table we present the SNR and
  correlation computed in the frequency domain between Kerr quadrupole
  and octopole waveforms for different EMRB orbits. The orbits are
  separated into two groups: highly-bursting (top) and slowly-bursting
  (bottom). The first five columns present information about the
  different orbits, including how many times they burst, their
  eccentricities and periods, which were chosen directly from the
  allowed EMRB phase space of~\citet{Rubbo:2006dv} and thus represent
  Milky Way sources. The sixth and seventh columns present the
  difference in SNR between a Newtonian and Kerr quadrupole waveform
  relative to the former using the entire data set and only one burst
  respectively. Similarly, the eighth and nineth columns show the
  correlation between the plus polarizations using the entire data set
  four and one burst respectively. Since the slowly-bursting sources
  burst only once, the sixth and eighth columns are redundant for
  these sources. All calculations assume the observer is located on
  the $z$-axis and random initial inclination angles.}
\end{deluxetable}
\end{center}

\clearpage 
\begin{center}
\begin{deluxetable}{c c c c c}
\tablewidth{0pt}
\tablecaption{\label{table_obs} Comparison between quadrupole and
  octopole waveforms.}
\tablehead{
\colhead{Angle [degrees]} & \colhead{$x_{\textrm{obs}}$ [kpc]} &
\colhead{$y_{\textrm{obs}}$ [kpc]} & \colhead{$z_{\textrm{obs}}$ [kpc]}
& \colhead{Amp.~diff.}} 
\startdata
$0$  & $0$  & $0$ & $8$ & $3.9 \%$ \\[1mm]
$20$ & $2.73$  & $0$ & $7.51$ & $3.3 \%$ \\[1mm]
$40$ & $5.14$  & $0$ & $6.12$ & $11.5 \%$ \\[1mm]
$60$ & $6.93$  & $0$ & $4$    & $15.2 \%$ \\[1mm]
$80$ & $7.88$  & $0$ & $1.39$ & $19.7 \%$ \\[1mm]
$100$ & $7.88$  & $0$ & $-1.39$ & $21.6 \%$ \\[1mm]
$120$ & $6.93$  & $0$ & $-4$ & $17.8 \%$ \\[1mm]
$140$ & $5.14$  & $0$ & $-6.13$ & $43.3 \%$ \\[1mm]
$160$ & $2.74$  & $0$ & $-7.52$ & $8.6 \%$ \\[1mm]
\enddata
\tablecomments{Here we present an approximate measure of the amplitude
  difference between the quadrupole and octopole waveforms. We
  concentrate only on the first burst of radiation and we normalize
  the difference to the maximum of the first peak of the quadrupole
  waveform. The difference is presented as a function of the observer
  location, which is always at a fixed radial distance of
  $r_{\textrm{obs}} = 8$ kpc, but rotated about the $\hat{y}$-axis on
  $\hat{x}-\hat{z}$ plane ($\theta$ is here the usual Euler polar
  angle.) }
\end{deluxetable}
\end{center}
%

\end{document}